# A Stimulus-Response Model for Explaining When Students Decide to Engage in a Science Task:

## Developing the TSMS Model Through Physics

Eva Cauet[1], Alexander Kauertz[1]

[1] Institute for Science Education, RPTU Kaiserslautern-Landau in Landau

## Author Note

Eva Cauet https://orcid.org/0000-0003-0440-1306
Alexander Kauertz https://orcid.org/0000-0003-1700-4190

Correspondence concerning this article should be addressed to:
Eva Cauet, RPTU Kaiserslautern-Landau, Fortstr.7, 76829 Landau. Email: eva.cauet@rptu.de

We have no conflict of interest to disclose.

## Acknowledgements

The initial version of this manuscript and the underlying TSMS model were developed and written by the authors without the use of generative AI. Following the first round of external peer review, the authors used ChatGPT (OpenAI; accessed 2025–2026) during the revision process to support language editing, the refinement of selected passages and transitions, feedback on clarity, coherence, organization, and potential redundancies, and the condensation and polishing of the Abstract. ChatGPT was also used to generate and refine draft formulations for selected revisions; to support the articulation of testable hypotheses and the semi-formal representation derived from the authors' TSMS model; and to suggest possible ways of visually representing the model based on the authors' initial design for Figure 1. The model's underlying theoretical ideas, core assumptions, conceptual structure, interpretation of the literature, and all substantive decisions remained under the authors' direction. Figure 1 was manually created, adapted, and finalised by the authors. All AI-assisted suggestions and text were critically reviewed, iteratively revised, and approved by the authors, who take full responsibility for the accuracy, originality, and integrity of the manuscript.

# A Stimulus-Response Model for Explaining When Students Decide to Engage in a Science Task: Developing the TSMS Model Through Physics

## Abstract

Science education research shows that students' reasoning can be shaped by contextual embedding, salient task cues, and interactions among cognitive, metacognitive, affective, and motivational processes. Yet how these processes interact in the predecisional phase of task processing, when students first interpret a task and decide how to engage, remains undertheorised. To address this gap, we introduce the Task-Specific Motivational Stimulus (TSMS) model, a theoretical account of students' early disciplinary engagement with science tasks developed through physics as a prototypical case. Drawing on expectancy-value theories, heuristic-analytic and metacognitive accounts of reasoning, and physics education research on task representation and resources-based reasoning, the model proposes that task cues trigger initial task typification through heuristic knowledge activation and integration, resulting in a potentially misleading first impression mental model of the task type (FIMM). Accompanying metacognitive feelings of rightness (FOR) and difficulty (FOD) shape task-related self-efficacy, value, and perceived cost. These evaluations are integrated into a task-specific motivational stimulus, the motivationally informed judgement of solvability (MJOS), which is assumed to influence the probability that students engage in disciplinary analytic reasoning to build a task representation or rely on unexamined initial interpretations. We justify the model's assumptions, specify and formalise it for physics tasks, derive testable hypotheses for future empirical validation, and discuss how it can organise physics education findings through a common explanatory mechanism. Although developed through physics, the TSMS model is proposed as a theoretical framework for investigating an empirically underexplored decision process underlying early engagement with science tasks that share similar characteristics.

.

## Introduction

Across science domains, educators and researchers are familiar with the challenge that students do not always apply previously learned concepts and scientific reasoning during science tests, and instead rely on non-scientific explanations despite knowing better (Potvin & Cyr, 2017; Sinatra et al., 2015). Findings from PISA 2003, for example, highlighted significant differences between German students' domain-unspecific analytical problem-solving competence and their competencies in mathematics and science, despite similar task demands (Prenzel et al., 2002). Fleischer et al. (2014) argued that students may not always apply their competencies in achievement test-situations and showed that students with low mathematical self-concept and high fear of mathematics performed notably worse when PISA mathematics and domain-unspecific problem-solving tasks were embedded in a 'maths test situation' by explicitly labelling them as mathematics tests. Because cognitive engagement is assumed to mediate the relationship between emotion and performance (Pekrun & Linnenbrink-Garcia, 2012), these findings suggest that test situations can reduce cognitive engagement when they activate unfavourable motivational dispositions and emotions. In physics tests, which we use in this paper as a prototypical case for developing our model, such variation may even occur from one task to another. Heckler (2011) and Kryjevskaia et al. (2014) showed students' answer patterns on multiple-choice physics tasks can be strongly influenced by subtle changes in task descriptions, leading to misconception-like responses in some tasks and correct responses in others.

A central role of tasks in science education is to facilitate the diagnosis of students' knowledge and competencies. In physics education, this diagnostic function is especially well documented (Fischer & Kauertz, 2021). However, task-based diagnoses can only allow valid conclusions on competencies when students seriously engage with these tasks (cf. Asseburg & Frey, 2013; Nicholls, 1984). Yet, Heckler (2011) showed that student answering patterns are strongly influenced by bottom-up processing: instead of engaging in physics-specific reasoning processes, students often rely on their initial thoughts when attempting to solve physics tasks. This might involve using experiential knowledge or applying the first physics concept that comes to mind, which may lead students to adopt an inappropriate line of reasoning, even when they draw on physics-specific concepts. But what determines whether students engage in disciplinary reasoning when working on a science task, and how can this question be studied in the prototypical case of physics tasks?

To answer this question, one must consider on which levels student engagement can be influenced. A task cannot be viewed independently of its surrounding context, as tasks are reconstructed by individuals and embedded in specific situations (Finkelstein, 2005).

Finkelstein (2005) identifies a situation frame which recognises the context in which a task is viewed, and a task formation frame which involves the interplay between the individual attempting to solve the task, the problem's storyline, and the underlying physics concept. Before students engage in the cognitive processes required to solve a task, the testing environment may prime an individual's knowledge, motivation, emotions, and beliefs pertaining to physics or physics tasks (ibid.). The empirical evidence mentioned above highlights the importance of considering interactions within these contextual frames when interpreting task performance and evaluating student competence.

Building on Finkelstein's distinction between the situation frame and the task formation frame, we first consider motivational theories that explain general antecedents of engagement within the situation frame. We then turn to the question of how engagement must be specified for science tasks and develop this argument in detail for physics tasks, which provide a particularly suitable case because they often require students to coordinate everyday experiences, formal representations, modelling assumptions, and abstract concepts (Löffler et al., 2018; Sabella & Redish, 2007).

**Antecedents of Student Engagement**

Low test-taking motivation can compromise the reliability and validity of test results (DeMars et al., 2013). This is particularly concerning in physics, where students often report unfavourable motivational dispositions, such as low interest and low competence beliefs (Romero-Abrio & Hurtado-Bermúdez, 2023; Steidtmann et al., 2023). However, even high test motivation does not necessarily translate into engagement.

Motivation, as defined by Cleary and Zimmerman (2001), refers to the will to use deep strategies during task solving, whereas engagement refers to the actual use of these strategies. Accordingly, Miller (2015) defines student engagement as the 'quantity and quality of mental resources directed at an object of thought […], as well as the emotional reactions and behaviours that lead to and enable using those resources' (p.33). Student engagement involves emotional, behavioural, and cognitive aspects that become increasingly interconnected when examining engagement at a more detailed level (Miller, 2015; Sinatra et al., 2015). Motivation is closely tied to each dimension of engagement (Sinatra et al., 2015) and is influenced by self-efficacy beliefs and value attributions (Eccles & Wigfield, 2020; Katstaller & Gniewosz, 2020). Indeed, antecedents of motivation are related to various indicators of engagement (see Greene & Miller, 1996 for an overview; Johnson & Sinatra, 2013). In line with this, Pekrun's (2006) control-value theory highlights the importance of self-efficacy beliefs, value attributions, and cost for emotional engagement, which can influence behavioural and cognitive aspects of engagement and, ultimately, performance.

While such motivational and emotional processes may be relevant across domains, Sinatra et al. (2015) call for more theoretical work on science-specific engagement, including 'more precise and differentiated measures of science engagement' and greater attention to how, 'motivational and emotional factors […] interact with how one chooses to engage with science content' (p.4). This suggests that understanding engagement in science tasks requires not only identifying general motivational antecedents, but also specifying how students engage with the cognitive demands of science content. In the next section, we develop this specification for physics tasks as the prototypical case considered in this paper.

**Task-Specific Engagement in Science Tasks: Physics as a Prototypical Case for the TSMS Model**

Pekrun and Linnenbrink-Garcia (2012) distinguish between pure cognitive engagement (attention and automatically activated memory processes), behavioural engagement (effort and persistence) and cognitive-behavioural engagement (strategy use and self-regulation). However, cognitive–behavioural engagement still requires further domain-specific specification. In line with Sinatra et al.'s (2015) call for more differentiated accounts of science engagement, we use physics tasks as a prototypical case to specify when cognitive–behavioural engagement becomes disciplinary reasoning. Physics tasks provide a particularly clear case of a broader challenge in science task engagement: students must coordinate disciplinary ideas with scientific practices already when trying to understand what a task requires (National Research Council, 2012). In physics, this challenge is especially visible because understanding a task often requires students to relate everyday experiences, formal or graphical representations, and modelling assumptions to the relevant physics model (e.g. Löffler et al., 2018; Sabella & Redish, 2007). This may involve strategies such as controlling variables, using analogies, coordinating symbolic expressions with models, or simplifying situations, as well as self-regulation processes such as inhibiting misleading interpretations or questioning familiar routines that cue inappropriate approaches.

We focus on the very beginning of task processing, where students decide how to approach a task. This is particularly relevant because, as discussed above, physics tasks often involve a tension between intuitive interpretations of phenomena and formal physics reasoning. The initial appraisal of a task therefore becomes consequential for how students engage with it: they must decide whether to rely on intuitive interpretations or to engage in physics-specific reasoning to examine whether their interpretation fits the task. In contrast, behavioural engagement in terms of effort and persistence primarily becomes relevant once students have already decided how to approach the task and started working on it. At this point, however, they may already have adopted a misguided line of reasoning.

Importantly, cognitive–behavioural engagement does not necessarily imply physics-specific reasoning. Students may use analogies without questioning their adequacy, try hard to recall familiar answers, or apply the first physics concept or formula that comes to mind (cf. M. K. Kaiser et al., 1986; Sabella & Redish, 2007). Thus, physics-specific cognitive–behavioural engagement at the beginning of task processing refers more narrowly to analytic reasoning that supports the construction of an adequate task representation by relating surface features to underlying physics concepts, principles, or models.

Insights from dual-process research help to describe why this distinction matters: while alternative perspectives on cognition exist, Evans' heuristic–analytic theory and Thompson's metacognitive framework of reasoning are particularly useful for describing when intuitive interpretations are maintained, elaborated, justified, or revised through analytic processing and accompanying metacognitive experiences (Evans, 2019; Thompson, 2009). Students may construct an initial interpretation of a task based on intuitive or heuristic processes, for example by comparing it to similar tasks from the past. Depending on how "right" this intuition feels, they may engage in analytic reasoning mainly to elaborate or justify the initial interpretation rather than to question it. Of course, it can be helpful to have a good intuition on how to solve a physics task (like experts would have), e.g., by intuitively activating relevant physics concepts or procedures. However, cognitively engaging with a task in a physics-specific way means not simply following a first idea without questioning it, but activating different concepts and procedures to decide which ones are appropriate for the task at hand (e.g., Heckler & Bogdan, 2018; Sabella & Redish, 2007). Neuroimaging studies show: Even experts exhibit strong inhibitory processes when working on physics tasks that trigger alternative conceptions (see Nenciovici et al., 2019 for an overview).

Physics-specific cognitive-behavioural engagement is therefore not simply a matter of "thinking hard" about a task. It involves employing domain-specific procedures such as breaking down complex problems, using mathematical representations, understanding physics terminology (e.g., force, energy), considering theories and core ideas, and applying structured problem-solving approaches (e.g., energy-based approaches). While thinking in a physics-oriented way certainly involves creativity, it typically involves modelling, analytic reasoning, adherence to causal logic, and the critical evaluation of intuitive impressions in light of formal physics principles. Everyday experiences can provide valuable background knowledge about physical phenomena, but interpretations derived from these experiences may not always align with the formalistic models used in physics. Although experiential student conceptions occur in almost all fields, they appear to be particularly prevalent and persistent in science (Sinatra et al., 2015). Research in physics education has documented the persistence of such conceptions extensively (see Chi et al., 2012 for an overview) and shows that intuitive and scientific conceptions can even coexist within a person and that their activation depends on contextual cues in the task description as well as students' executive functions such as inhibition or set shifting (Potvin, 2013). Many physics tasks are contextualised through everyday situations, yet include simplifications and idealisations, such as neglecting air resistance. In such tasks, physics-specific reasoning may require student to actively inhibit everyday interpretations and instead relate task cues to the relevant physics model, which could explain the persistence of student conceptions in physics (Mason & Zaccoletti, 2021).

While research on student engagement and its motivational antecedents and research on reasoning have provided important insights, these approaches remain largely separate. Engagement research explains why motivational and emotional factors matter for learning and performance, but often remains located at the level of the situation frame and does not specify how students decide whether to engage in disciplinary reasoning for a particular task. Reasoning research explains how intuitive and analytic processes operate in the task formation frame and shows that metacognitive experiences can influence whether initial interpretations are accepted or reconsidered. However, it often does not specify how such experiences become motivationally relevant for students' task-specific decision to engage. Little theoretical work has focused on the decision students make at the very beginning of task processing. Yet, as argued above, this decision point is crucial in science tasks because students may either proceed on the basis of an initial interpretation or engage in disciplinary analytic reasoning to examine whether this interpretation fits the task. To explain how this decision is made, we need a model that links these theoretical perspectives at the interface between the situation frame and the task formation frame.

We address this gap by proposing a theoretical model of task-specific engagement in science tasks, developed through the prototypical case of physics task processing. The model responds to Sinatra et al.'s (2015) call for a more domain-specific account of science engagement by explaining how students' initial task interpretation, metacognitive experiences, and motivational evaluations interact in this decision-making process at the very beginning of task processing. Therefore, our model focuses on what we later introduce as the *predecisional phase of task processing* and aims to explain when students engage in disciplinary analytic reasoning rather than proceed on the basis of an unexamined initial interpretation.

Disentangling the complex interplay between mental task representations, contextual cues, and motivational antecedents of student engagement may help to better understand how students engage with science tasks. Such insights may also inform the design of tasks or scaffolds that support science-specific engagement. While this would certainly not guarantee successful task solving, it may help improve the validity of task-based diagnoses by increasing the likelihood of science-specific engagement with these tasks.

## The Task-Specific Motivational-Stimulus Model

Symonds et al. (2019) state that momentary engagement in a specific context, such as taking a science test or working on a specific physics task, always needs a trigger. We therefore describe the decision to engage with a specific science test task as part of a stimulus-response scheme, with student engagement being the response to a motivational stimulus created by the task. Building on the considerations outlined above, we aim to explain how such a motivational stimulus emerges when students encounter a science task and how it influences their decision to engage in disciplinary reasoning.

In this paper, we present the *Task-Specific Motivational Stimulus Model* (TSMS model) as a theoretical model of task-specific engagement in science tasks, developed through the prototypical case of physics task processing. The model explains how students' initial task interpretations, metacognitive experiences, and motivational evaluations interact at the beginning of task processing and shape whether students engage in disciplinary analytic reasoning.

The TSMS model integrates insights from research on student engagement and its motivational antecedents, especially expectancy–value and control-value approaches, with heuristic–analytic theories of reasoning. To clarify how these perspectives contribute to specific components of the model, Table 1 provides an orienting overview of the model's core assumptions, their theoretical grounding, and their incorporation into the model, while Figure 1 visualises the model's process logic. Because the present paper develops the TSMS model through the prototypical case of physics tasks, Table 1 and Figure 1 present the physics-specific specification of the model. The first three assumptions embed the TSMS model in the broader situation frame and describe how the physics test situation may shape initial task processing. Assumptions 4–6 specify the model's core mechanism: during initial task typification, students construct a first impression mental model of the task type (FIMM), through heuristic processes whose accompanying metacognitive experiences, a feeling of rightness (FOR) and a feeling difficulty (FOD), specify task-related motivational evaluations integrated in the motivationally informed judgement of solvability (MJOS) which is assumed to influence whether students engage in physics-specific reasoning or proceed on the basis of an unexamined initial interpretation. The following sections justify the assumptions and constructs in detail and specify the assumed relations between motivational, cognitive, and metacognitive processes in early task engagement. In line with calls for more explicit and testable theory building (e.g. Fried, 2020) the model is then summarised in a semi-formal representation, from which testable hypotheses are derived before the model's scope, empirical plausibility, theoretical contribution, and implications are discussed.

*Table 1*
*Overview of the Model Assumptions, Underlying Theoretical Background, and Their Incorporation into the TSMS Model.*

| | **Assumption** | **Theoretical Background** | **Incorporation into the TSMS model** (the bolded terms link the content with Figure 1) |
|---|---|---|---|
| **Contextual embedding: Assumptions in the situation frame** | (1) Expectancy-value evaluations, based on physics-specific self-efficacy beliefs and value/cost attributions activated by a physics test situation, constitute a situational motivational stimulus which triggers emotional engagement within the situation frame. | • Control-value theory of achievement emotions (Pekrun, 2006; Pekrun & Linnenbrink-Garcia, 2012)<br>• Situated expectancy-value theory for achievement motivation (Eccles & Wigfield, 2020) | • The test situation activates **physics-specific self-efficacy beliefs** and **value** attributions (**intrinsic** value, **extrinsic** value, and **cost**) regarding working on physics tasks in general.<br>• The resulting expectancy-value evaluations constitute a **situational motivational stimulus** that shapes students' **achievement emotions** within the **situation frame** (e.g., anxiety, hope, hopelessness).<br>• Which emotion is triggered depends on students' achievement goals and focus of value attribution. |
| | (2) Emotional engagement can influence cognitive engagement in text-comprehension processes during first task reading. | • Process, Emotion, and Task (PET) framework (Bohn-Gettler, 2019) | • **Emotional engagement** links the situational motivational stimulus to **cognitive engagement** during task reading by shaping attention, cognitive resources, and knowledge activation or integration, depending on the emotion, comprehension process, and task demands.<br>• In this way, **achievement emotions** can influence the cognitive processes through which students construct an initial task representation during first task reading, as specified in Assumption n°5. |
| | (3) During first task reading, students enter the predecisional phase of task processing, in which the situational motivational stimulus can begin to be specified for the task at hand. | • Rubicon model of goal-oriented action phases (Heckhausen & Gollwitzer, 1987) | • The **predecisional phase** begins when students enter the task formation frame during first task reading and make an initial decision about how to engage with the task content.<br>• During this phase, broad physics-related self-efficacy beliefs, value attributions, and perceived costs may begin to shift toward evaluations related to the task at hand.<br>• This early specification depends on how students initially interpret the task and on the **metacognitive experiences accompanying** this initial interpretation; their role in shaping task-related motivational evaluations is specified in Assumption n°6. |

*Note.* Continued on the next page

*Table 1 (continued)*
*Overview of the Model Assumptions, Underlying Theoretical Background, and Their Incorporation into the TSMS Model.*

| | Assumption | Theoretical Background | Incorporation into the TSMS model (the bolded terms link the content with Figure 1) |
|---|---|---|---|
| | (4) Initial task typification provides the task-type reference point for specifying the motivational stimulus. | • Problem Solving in Physics (Löffler et al., 2018; Sabella & Redish, 2007; Tuminaro & Redish, 2007)<br>• Expert-Novice Research in Physics (Chi et al., 1981; Jong & Ferguson-Hessler, 1991; Schoenfeld & Herrmann, 1982) | • In physics tasks, the relevant task type and solution approach are often not directly visible from surface features; students therefore form an initial typification of what kind of task they are facing and what it appears to require. This early typification can strongly shape subsequent reasoning<br>• If **initial task typification** appears sufficiently credible, task-type-specific self-efficacy beliefs, value attributions, and cost attributions may become available and provide the first task-type reference point for specifying the situational motivational stimulus. If typification remains unclear or lacks credibility, broader physics-related motivational representations should remain more influential. |
| **Core TSMS mechanism: Assumptions in the task formation frame** | (5) During initial task typification, students construct a first impression mental model of the task type through heuristic processes. | • Heuristic-analytic theory of reasoning (Evans, 2006, 2019)<br>• Resources Model (e.g. Hammer et al., 2005) | • **Initial task typification** begins when students process **task cues** through heuristic activation and integration of **resources**, such as intuitive or conceptual knowledge elements, epistemological beliefs, strategies, task schemata, and affectively charged experiences, into an emerging **first impression mental model of the task type (FIMM)**.<br>• **Cognitive engagement** shapes how broadly the resources are activated, and which elements are maintained and integrated. Activated **beliefs** about physics tasks or heuristics as the availability heuristic influence which cues and resources appear relevant; as reading continues, the emerging **FIMM** recursively guides further activation and integration.<br>• The **FIMM** is the first provisionally coherent configuration of **activated resources** that is sufficiently stable to guide subsequent task processing and provides an initial idea of how to approach the task. It is not yet validated and may lead to an inappropriate line of reasoning. |
| | (6) Metacognitive feelings that accompany these processes shape task-related motivational evaluations and thereby specify the task-specific motivational stimulus for physics-specific engagement. | • Metacognitive framework of reasoning (Thompson, 2009) | • A metacognitive **feeling of rightness (FOR)** and **feeling of difficulty (FOD)** accompany the heuristic construction of the **FIMM** and specify task-related **self-efficacy**, **value**, and perceived **costs**.<br>• **FOR** indicates how credible the emerging **FIMM** feels. Higher **FOR** should increase the influence of task-type-specific self-efficacy beliefs, value attributions, and cost attributions on task-related motivational evaluations, whereas lower **FOR** should leave broader physics-related self-efficacy beliefs, value and cost attributions more influential. **FOD** provides an initial signal of difficulty. Higher **FOD** should increase task-specific perceived cost and reduce task-specific net value (i.e. value adjusted for cost).<br>• The resulting task-related motivational evaluations are integrated in the **motivationally informed judgment of solvability (MJOS)**, which represents the **task-specific motivational stimulus**. **MJOS** predicts the probability that students engage in **physics-specific reasoning processes for building a task representation** by questioning their **FIMM** and, if necessary, revising it – with engagement expected to be most likely at intermediate **MJOS** values. |

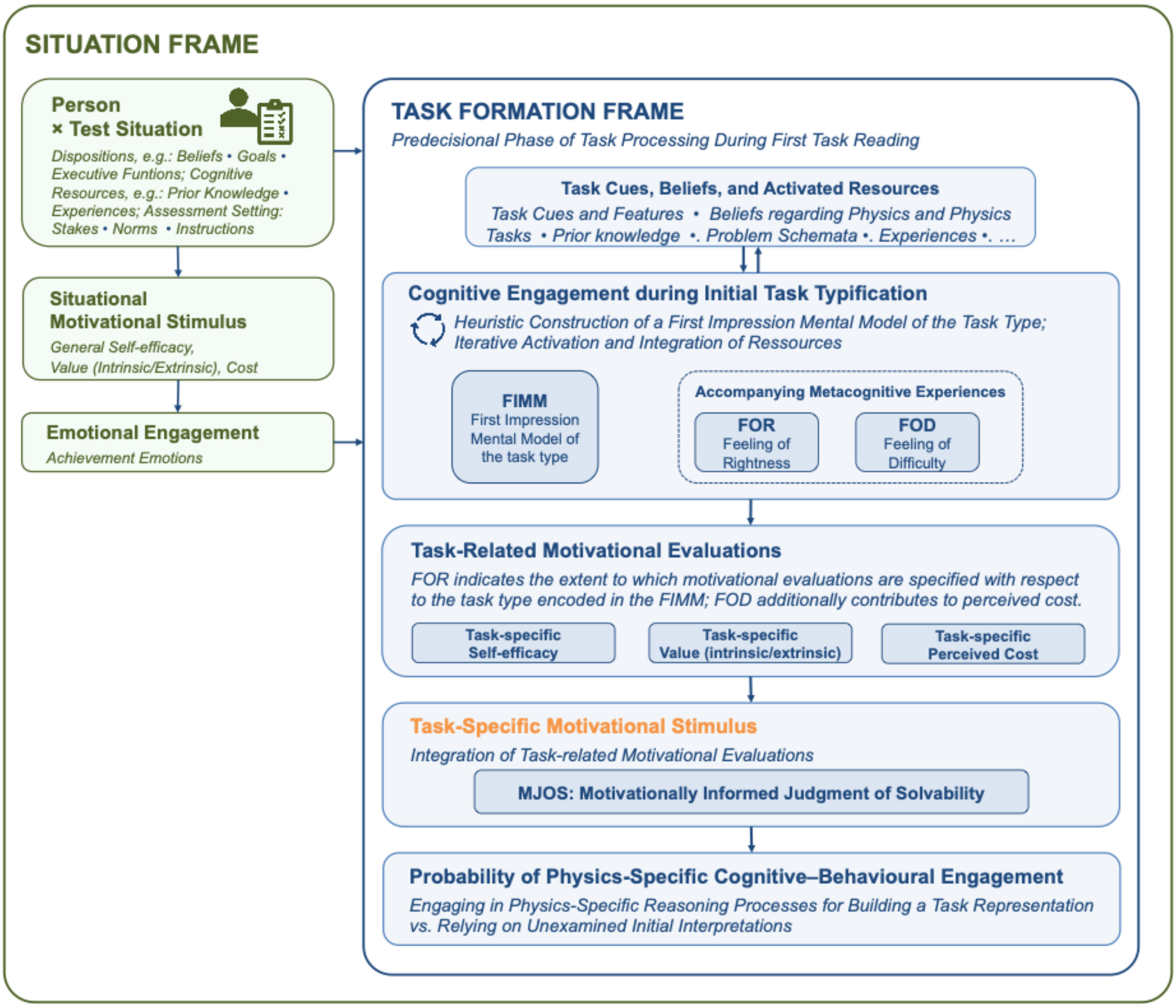

*Figure 1: Process Logic of the Task-Specific Motivational Stimulus (TSMS) Model*

## Basic Assumptions of the TSMS Model

In the following, we will theoretically justify the assumptions in Table 1, present empirical findings to support them, and explain how we have incorporated these findings into the TSMS model.

### Contextual Embedding: Assumptions in the Situation Frame

The TSMS model is based on some domain-unspecific assumptions that address how engagement within the situation frame is built and how it influences motivational and cognitive processes in the task formation frame:

- Expectancy-value evaluations, based on physics related self-efficacy beliefs and value/cost attributions activated by a physics test situation, constitute a situational motivational stimulus which triggers emotional engagement within the situation frame (Assumption n°1).
- Emotional engagement within the situation frame can influence cognitive engagement in text-comprehension processes during task reading (Assumption n°2).
- During first task reading, students enter the predecisional phase of task processing, in which the situational motivational stimulus can begin to be specified for the task at hand. (Assumption n°3).

Expectancy-value models have a long tradition in motivation theory (e.g., see Atkinson, 1957). They conceptualise motivation as the output of expectancy-value evaluations based on students' subjective expectation of success ("Am I able to solve this task?"), usually operationalised as self-efficacy beliefs (Eccles & Wigfield, 2020), and their subjective attribution of value to a task ("Why should I solve this task"), which characterises the individually perceived importance attributed to success (or avoiding failure) in the test situation, i.e., intrinsic value (e.g., interest or attainment value), extrinsic value (e.g., utility value or reward), and cost associated with working on or solving a physics task. Prominent examples are the situated expectancy-value theory for achievement motivation (Eccles & Wigfield, 2020) or the control-value theory for achievement emotions (Pekrun, 2006). According to Pekrun (2006) expectancy-value evaluations determine emotional engagement within the situation frame (Assumption n°1). The focus of value attribution (expectation of success or fear of failure) depends significantly on the underlying achievement goals and determines which emotion is triggered (Pekrun et al., 2009; Pekrun & Stephens, 2010). The intensity of the triggered emotions is proportional to the level of value attribution, but can vary positively, negatively, or curvilinearly with expectations of control depending on the type of triggered emotion. Pekrun and Stephens (2010) classify achievement emotions regarding their valence (negative vs. positive) and their potential for activation which determines their effects on engagement and performance (see Pekrun & Linnenbrink-Garcia, 2012 for an overview of empirical studies).

According to the Process, Emotion, and Task (PET) framework (Bohn-Gettler, 2019), emotional engagement influences students' text comprehension processes during task reading, particularly at the situation model level, where connections are made between the current sentence and prior knowledge. However, these effects are not determined by emotional valence alone, but depend on the specific emotion, comprehension process, and task demands. Positive emotions may support assimilative processing, whereas negative emotions may support accommodative or detail-oriented processing when revision of an existing representation is required (Bohn-Gettler, 2019; Fiedler & Beier, 2014). Thus, emotional engagement may influence which knowledge elements are activated, maintained, integrated, or validated during task reading. Because emotional engagement can change when students encounter a specific task (e.g., when a task appears easier than feared), emotional engagement in the situation frame will predominantly influence students' domain-unspecific cognitive engagement when reading the task for the first time in the task formation frame (Assumption n°2).

Indeed, theoretical and empirical work suggests that antecedents of engagement can be specified for the task at hand. Situated expectancy–value theory (Eccles & Wigfield, 2020) conceptualises expectancy–value evaluations as context-sensitive and subject to change depending on situational and task-specific characteristics. In line with this perspective, empirical research on self-efficacy suggests that efficacy judgements can be differentiated at narrower levels of specificity: in science education, task-specific self-efficacy for scientific reasoning and science content knowledge can be distinguished from general academic self-efficacy (Nyberg et al., 2022); in mathematics education, self-efficacy judgements have been shown to be sensitive to task characteristics such as difficulty, representational format, and context (e.g., Chen & Zimmerman, 2007; Siefer et al., 2021). Findings on the task specificity of value components are less consistent. While Schukajlow et al. (2012) did not find differences in students' values across different types of mathematical problems, Krawitz and Schukajlow (2018) reported that value-related beliefs can vary across problem types. In addition, Efklides et al. (2006) showed that expected costs associated with working on a task can be updated during task formation, with perceived item difficulty influencing prospective estimates of required effort. Taken together, these findings suggest that motivational evaluations can become task-related, although the conditions and mechanisms of such specification remain insufficiently understood and may vary, for example depending on how readily task characteristics can be interpreted.

The TSMS model addresses this gap by proposing that such specification begins during first task reading (Assumption n°3) and is shaped by the initial typification of the physics task and the metacognitive experiences accompanying it. This mechanism can account for the differing empirical findings reported above (see discussion).

A successful solving process requires building a goal intention to at least try to solve the task. This goal intention is the basis for the decision to engage in seriously processing a task. The Rubicon model of goal-orientated action phases from Heckhausen and Gollwitzer (1987) provides a useful framework for describing 'the transition from a motivational state of deliberation to the volitional state of implementation' (p. 103) in a so called *predecisional* and a *postdecisional phase* of task processing. Empirical studies suggest that the predecisional phase is characterised by deliberate mindsets and motivational task-specific expectancy–value evaluations (e.g., Fujita et al., 2007; Puca & Schmalt, 2001).

In the TSMS model, these theoretical perspectives are incorporated as follows. Expectancy–value and control-value approaches explain how the physics test situation activates self-efficacy, value, and cost evaluations that constitute the situational motivational stimulus and shape emotional engagement in the situation frame. The PET framework explains how this emotional engagement can influence early text-comprehension processes during first task reading, especially knowledge activation and integration. The Rubicon model justifies the focus of the TSMS model on the predecisional phase, before students have committed to a way of approaching the task. The reading and first understanding of the task constitute a key moment for the decision to engage with it in a physics-specific way, because this is where the situational motivational stimulus can begin to become specified for the task at hand and where the initial task representation starts to guide the direction of subsequent reasoning. Once this initial decision is made, the subsequent task-solving process is shaped by the initial task representation and the associated motivational state, although it may still be interrupted, redirected, or terminated during later processing.

The next section therefore turns to the physics-specific processes through which this initial task interpretation is formed and through which the situational motivational stimulus becomes specified for the task at hand.

## Core TSMS Mechanism: Assumptions in the Task Formation Frame

Having embedded the TSMS model in domain-unspecific accounts of motivational, emotional, and early cognitive engagement, we now turn to the physics-specific person–task interaction during first task reading that may shape whether students engage in physics-specific reasoning to validate or revise their initial task interpretation. To explain how motivational and emotional processes influence cognitive engagement in the task formation frame, we have integrated models from domain-unspecific research on engagement. We will now explore triggers in the person-task interaction that prompt students to employ physics-specific reasoning processes to validate and adapt their first impression mental model of the task type to a more suitable situation model when needed. We will first argue that during the first reading of a physics task, students form an initial, not yet validated typification of the physics task, which provides a reference point for specifying the situational motivational stimulus for the task at hand (Assumption n°4). We then specify the cognitive output of this initial task typification: the first impression mental model of the task type (FIMM), which students construct through heuristic processes (Assumption n°5). Finally, we will argue that the metacognitive feelings and judgements accompanying the construction of the FIMM – namely a feeling of rightness and a feeling of difficulty – shape the task-related motivational evaluations and thereby specify the task-specific motivational stimulus for physics-specific engagement (Assumption n°6). Although these assumptions partly draw on domain-unspecific models, they are specified here for physics tasks to explain how students' initial task interpretations and accompanying metacognitive experiences become motivationally relevant for physics-specific engagement – and therewith for the decision to engage in physics-specific reasoning processes to question an initial task interpretation rather than proceed with it unexamined.

***Initial task typification provides the task-type reference point for specifying the motivational stimulus. (Assumption n°4).***

In physics tasks, the relevant task type and solution approach are often not directly visible from surface features. Contextualised tasks may require students to infer the relevant physics content from problem situations embedded in everyday contexts or socio-scientific issues (OECD, 2019), whereas more formal or advanced physics tasks may require students to interpret equations, diagrams, graphs, or symbolic representations in relation to underlying models and principles (e.g. Sherin, 2001; Tuminaro & Redish, 2007). In both cases, surface features can cue misleading initial typifications: everyday experiences may suggest inappropriate intuitive interpretations, formal features may trigger familiar but inadequate procedures, such as searching for a formula or assuming that calculation is required, and directly activated but inappropriate knowledge structures – as well as beliefs about physics tasks – can inhibit activation of useful structures (Bing & Redish, 2009; Gupta & Elby, 2011; Sabella & Redish, 2007). To understand what such tasks require, students need to relate surface features to underlying concepts, principles, and models at the deep-structure level (Löffler et al., 2018).

Establishing these links requires strategy use and the effortful construction of an adequate situation model of the task. Novices and experts typify such tasks based on the strategy they consider useful for that purpose. However, novice strategies are bottom-up and often linked to surface elements, whereas experts use generalised and model-related strategies, such as identifying relevant concepts, principles, and problem schemata (Chi et al., 1981; Jong & Ferguson-Hessler, 1991; Schoenfeld & Herrmann, 1982). The importance of identifying such underlying problem schemata has been shown by Binder et al. (2019): in physics, students' ability to identify underlying problem schemata in sorting tasks predicted first-year academic achievement, whereas the corresponding application-of-knowledge measure was not predictive in biology. Moreover, physics tasks often activate everyday experiences that may conflict with the simplified or idealised models used in physics. In such cases, students may need to inhibit intuitive interpretations in order to construct an adequate task representation (Mason & Zaccoletti, 2021). Findings from Kryjevskaia et al. (2014) further suggest that even uncontextualised physics tasks can cue misleading interpretations that guide subsequent reasoning in an inappropriate direction. Taken together, these findings indicate that students' early orientation toward a physics task can strongly shape subsequent reasoning.

In the TSMS model, we describe this early, provisional orientation as initial task typification: when students first read a task, they form an initial impression of what kind of task they are facing and what it appears to require. The recognised task type may refer to conceptual aspects, such as energy or force; content areas, such as astronomy or motion analysis; surface aspects, such as long reading time, numerical information, or the need to draw a sketch; or affectively charged task experiences, such as tasks perceived as interesting, boring, or threatening. It therefore provides the task-type reference point for specifying the situational motivational stimulus for the task at hand. If the task appears familiar, task-type-specific self-efficacy beliefs, value and cost attributions may become available. For example, a student who recognises a task as an interesting real-world problem on energy use may activate high self-efficacy beliefs and positive value, but also task-type-specific cost expectations if the task appears easy but yet time-consuming because the same familiar calculation has to be repeated several times. If the task type remains unclear, broader physics-related beliefs and attributions should remain more influential. However, initial task typification does not by itself determine the task-specific motivational stimulus. Rather, the metacognitive experiences accompanying this initial typification influence the extent to which task-type-specific or more general representations dominate task-related motivational evaluations, as specified in Assumption n°6.

We now turn to the cognitive processes underlying this initial typification and argue that students construct a first impression mental model of the task type through heuristic processes, which represents the cognitive output of initial task typification.

***During initial task typification, students construct a first impression mental model of the task type through heuristic processes. (Assumption n°5).***

To describe these cognitive processes, we draw on dual-process theories because they explicitly address the relation between intuitive interpretations and analytic processing (Evans & Stanovich, 2013; Pennycook, 2018). We focus on Evans' extended heuristic–analytic theory (2019) because it is particularly suited to physics tasks in two respects. First, it is compatible with newer dual-process accounts suggesting that intuitive responses can include learned or logical intuitions rather than only naïve misconceptions (de Neys & Pennycook, 2019), which is important for describing expert-like reasoning in physics tasks. Second, it assumes that analytic processing can serve different functions: it may elaborate or justify an intuitive response, but it may also question and revise it when there is reason to do so. This distinction is useful for physics tasks because, as argued above, cognitive–behavioural engagement may involve analytic activity, such as applying a formula or elaborating an analogy, without necessarily involving physics-specific reasoning that challenges the initial task interpretation.

The extended heuristic-analytic theory of reasoning by Evans (2006, 2019) posits that intuitive heuristic processes generate selective mental models of the situation described in a task, contextualising it in a way that matches the current goal of the reasoner. Deliberate analytic processes may then elaborate, justify, or revise this initial mental model for task representation. Evans outlines three principles on which heuristic processes operate. The singularity principle posits that only one model can be processed at a time. The first impression mental model is built based on prior experiences and prior knowledge activated by cues in the task description. Logical intuitions, as described by De Neys and Pennycook (2019), may also play a role in its construction. According to the relevance principle, the first impression mental model is pragmatically inferred to be the most relevant and credible model for task representation. Thus, it not only describes a structural analogue of the situation outlined in the task, but also represents states of belief and knowledge (Evans, 2006, 2019). The satisficing principle suggests that analytic processing tends to accept the first impression mental model and only revises it when there is good reason to do so. If no such reason is perceived, analytic processes may be used to elaborate or justify the initial model (confirmation bias; Nickerson, 1998) or to generate direct inferences from it (Evans, 2006, 2019). Whether analytic processes are initiated to revise the first impression mental model depends on various factors, including motivational factors such as the importance of the decision, feeling of rightness (Thompson & Morsanyi, 2012), and thinking disposition (cf. Kryjevskaia & Grosz, 2020; Thompson et al., 2011), situational factors such as time available and context (cf. Evans & Curtis-Holmes, 2005; Heckler et al., 2010), and cognitive resources such as working memory capacity (Neys, 2006).

While Evans (2006, 2019) provides vast empirical evidence for the extended heuristic-analytic theory of reasoning from the reasoning and decision-making literature, Kryjevskaia and colleagues explicitly tested its applicability to students' reasoning processes when working on physics tasks (Kryjevskaia & Grosz, 2020; Kryjevskaia et al., 2014; Speirs et al., 2021). Kryjevskaia et al. (2014) showed that physics students sometimes used relevant physics knowledge to justify incorrect responses cued by an inappropriate initial interpretation rather than to revise that interpretation. This highlights how strongly the first impression mental model of a task can influence subsequent reasoning. Heckler (2011) summarises further evidence for the relevance of heuristic processes for misconception-like answers to physics questions.

In the next paragraph, we describe the construction of the first impression mental model of a task from a physics-specific perspective.

**The first impression mental model of the task type (FIMM) for physics tasks.** When students read a physics task for the first time, they form an initial impression of what they are expected to do and which kind of knowledge may be relevant for solving the task. We refer to this initial, not yet validated representation as the first impression mental model of the task type (FIMM). The FIMM is based on students' individual perception of the situation described in the task, which is informed by prior knowledge, prior experiences, and beliefs (e.g., Johnson-Laird, 1983; Nersessian, 1999; Vosgerau, 2006).

From a constructivist perspective and in accordance with models on text-comprehension (Bohn-Gettler, 2019), surface features of the task — such as physics terms, symbols, representations, contextual elements, or task format — act as cues that activate memories, experiences, emotions, and knowledge elements related to the situation described in the task. These resources can (but must not) be linked to conceptual knowledge elements or strategies for decoding the deep structure of the task. Normative or disciplinary knowledge may also be directly activated when it has been sufficiently practised to become available to heuristic processing (Stanovich, 2018). This may include symbolic forms, epistemic games, or familiar problem-solving strategies that can support reasoning but may also bias the FIMM when applied without considering the task's deep structure (e.g. Heckler, 2011; Sherin, 2001; Tuminaro & Redish, 2007). Thus, the FIMM can include both naïve resources and learned or expert-like intuitions. The FIMM is constructed through heuristic processes. From the *resources* perspective in physics education (e.g. Hammer et al., 2005), it can be understood as the currently most plausible configuration of activated resources, including intuitive knowledge elements (such as p-prims like 'more is more'), epistemological beliefs, problem-solving strategies, and affectively charged experiences (diSessa, 1993, 2018; Hammer et al., 2005; Sabella & Redish, 2007). Heuristics such as the availability heuristic – according to which the most easily accessible information is interpreted as most relevant (Tversky & Kahneman, 1973) – may influence which elements are prioritised. If students have experience with similar physics tasks, they may also activate task schemata or coordination classes from long-term memory (diSessa, 2018). For experts, these schemata may include procedural and conditional knowledge about appropriate problem-solving strategies (Chi et al., 1981), for novices, they may consist of more loosely connected declarative knowledge elements or more naïve strategies such as searching for formulas based on given variables.

Experiences and beliefs can also serve as reference categories for typifying a task. Kaiser et al. (1986), for example, found that students used reasoning by analogy as a default strategy for motion tasks and did not analyse the situation in formal physical terms when it could be mapped onto familiar experiences. Similarly, beliefs about physics tasks can shape the FIMM: students who believe that physics tasks typically require calculations may focus on numerical information and initiate inappropriate solution strategies, even when such information is irrelevant. Indeed, Gupta and Elby (2011) showed that students' epistemological beliefs about physics tasks and equations influenced their mathematical sense-making in physics tasks.

Furthermore, if a task shares salient surface structure features with other familiar tasks, it may be categorised as being of the same task type without further analysis (representativeness heuristic; Kahneman & Tversky, 1972). Kryjevskaia et al. (2014) provide an example where the representativeness and availability heuristics lead to an incorrect FIMM: half of the students who correctly applied their knowledge about the relationship between charge, capacity, and potential difference in capacitors in screening items failed to apply the same knowledge in a target item that cued intuitive reasoning based on the concept of conservation – which is a very well known concept and therefore highly available for students. Similar mechanisms are known from mathematics, where keyword strategies can lead students to translate words such as “more” or “less” directly into mathematical operations rather than constructing an adequate mental model of the task (Reusser & Stebler, 1997).

As students continue reading, the FIMM is assumed to be gradually constructed sentence by sentence. Each new surface feature and each newly activated resource is interpreted against the background of the already emerging FIMM. In this sense, the heuristic construction of the FIMM reflects cognitive engagement during text comprehension: students allocate attention, activate prior knowledge, and integrate selected elements into an emerging interpretation. Elements that fit this emerging structure may be integrated, whereas elements that do not fit may be ignored or read over. Importantly, this first provisionally coherent state need not emerge only after the whole task has been read, rather it may stabilise before reading is completed and may already guide subsequent task processing. We use the term FIMM to refer to the first provisionally coherent state of this construction process that is sufficiently stable to guide subsequent task processing: it represents the student's first impression of what kind of task this is and what it appears to require.

The next section builds on research on metacognitive feelings and judgements to argue that the cognitive output of initial task typification, the FIMM, is accompanied by a feeling of rightness and a feeling of difficulty, and to discuss how these feelings shape the task-specific motivational stimulus.

***The metacognitive feelings and judgments that accompany these processes shape the task-related motivational evaluations and thereby specify the task-specific motivational stimulus for physics-specific engagement. (Assumption n°6).***

In her metacognitive framework of reasoning, Thompson (2009) identified two metacognitive processes that influence whether analytic processing is initiated: the feeling of rightness (FOR), which accompanies an initial heuristic response, and the judgement of solvability (JOS) for the task. If FOR is high, the initial response is more likely to be accepted, and analytic processing is, if initiated at all, mainly used to justify it. However, if FOR is low, analytic processing is more likely to be initiated to revise the initial response. Thompson et al. (2011) showed for different types of domain-unspecific reasoning tasks if FOR was low that participants invested more time in rethinking their intuitive responses given under time constraints and more often changed them when given unlimited time. According to Thompson (2009), it depends on the judgement of solvability (JOS) if a reasoner reformulates the initial mental model when FOR is low. This assumption is based on the well-proven cognitive miser hypothesis – people only invest time and effort when probability of success is high enough (see Stanovich, 2009 for empirical evidence) – which is consistent with the expectancy-value models discussed before highlighting their shared assumption that the allocation of cognitive effort depends on perceived costs, benefits, and likelihood of success. Thompson (2009) further assumes that motivation and an initial feeling of difficulty (FOD; Efklides et al., 1999) influence the strength of the JOS. Lauterman and Ackerman (2019) showed that the JOS (yes or no), measured after four seconds of exposure to a matrix task, predicted the time spent solving the task.

In the next paragraphs we describe how we incorporate these theoretical ideas and findings into the TSMS model and provide further empirical to support this incorporation.

**Feeling of rightness (FOR) and feeling of difficulty (FOD).** We identify the FIMM as the heuristic response to the question what students thinks they are expected to do for solving a physics task (e.g., finding an analogy from everyday life which can be applied to the task, applying a p-prim, calculating something, applying energy conservation). Following Thompson's (2009) metacognitive framework of reasoning, we assume that, for physics tasks, the heuristic construction of the FIMM is accompanied by a feeling of rightness (FOR). But what does 'rightness' mean in this context? In our notion, the feeling of rightness is the feeling that the intention of the task has been correctly understood and indicates how credible the initial task typification feels and therefore how strongly self-efficacy, value and cost can be specified with reference to the identified task type. Thus, it denotes the impression of 'I know what I should do' – which does not necessarily mean 'I think I am able to do that'.

While FOR directly guides the acceptance of an initial response in Thompson's (2009) framework, its role in the TSMS model is embedded in a broader motivational evaluation captured by the motivationally informed judgement of solvability (MJOS). Accordingly, FOR informs the extent to which task-specific expectancy–value evaluations rely on task-type-specific rather than more general physics-related evaluations. High FOR enables the activation of task-type-specific self-efficacy beliefs, value and cost attributions, whereas low FOR leads to a stronger reliance on more general physics-related self-efficacy, value and cost – of which the former may only be slightly modified because of the experienced failure to identify what one is expected to do for solving the task.

Moreover, and in line with empirical findings (e.g. Efklides et al., 2006), the perceived cost associated with performing the task is additionally influenced by the feeling of difficulty (FOD), which accompanies the heuristic construction of the FIMM and reflects an initial estimate of task difficulty. Lower FOD implies lower perceived costs and thus requires less motivation to initiate task solving, whereas higher FOD implies higher perceived costs and requires stronger motivation (Barron & Hulleman, 2015). Extreme combinations may lead to overconfidence or disengagement.

While FOR and FOD shape students' initial evaluation of the task, the decision to engage in further processing is not determined by these feelings alone. The TSMS model assumes that it results from a broader integration of metacognitive and motivational evaluations. Specifically, the initial interpretation of the task (FIMM), the associated feeling of rightness (FOR), and the perceived difficulty (FOD) jointly contribute to the formation of a task-specific judgement that reflects students' perceived chances of success, perceived value, and perceived costs of engaging with the task. This view is consistent with research on self-regulated learning suggesting that metacognitive experiences and motivational evaluations jointly regulate effort and task engagement (e.g. Efklides, 2011).

**Motivationally informed judgement of solvability (MJOS): a model-based specification of the task-specific motivational stimulus.** Our conceptualisation of the task-specific motivational stimulus builds on Thompson's judgment of solvability (JOS), which was originally introduced as a metacognitive judgement about whether a problem appears solvable and worth further processing. However, later empirical operationalisations have often captured JOS more narrowly as a judgement of whether a task is solvable or not (e.g. Lauterman & Ackerman, 2019; Topolinski et al., 2016). We retain the broader theoretical idea but extent it by explicitly incorporating motivational considerations derived from expectancy–value theory. We therefore refer to it as the motivationally informed judgement of solvability (MJOS). MJOS is conceived as an integrative judgement of whether engaging with a task appears worthwhile and manageable: it reflects students' subjective impression of their ability to solve the task, taking into account the perceived cost of working on the task and the willingness to invest effort based on the value attributed to solving it in the given test situation. Thus, MJOS represents a task-specific instantiation of expectancy–value evaluations and provides a basis for future empirical operationalisation and testing of the model.

In the TSMS model, MJOS emerges from the interaction of metacognitive experiences (FOR, FOD) and motivational evaluations during initial task typification. At the end of the predecisional phase, students have constructed a FIMM, which provides an initial impression of what the task requires; a FOR, which indicates how right this FIMM feels and informs the source of task-related self-efficacy, value and cost; and a FOD, which additionally contributes to perceived cost for the task at hand. Together, FOR and FOD contribute to MJOS by shaping how students evaluate their chances of success, the perceived cost of engaging with the task, and the value of doing so.

MJOS is assumed to influence whether students engage in physics-specific reasoning processes for updating their FIMM or rely on their (potentially biased) initial interpretation of the task. Building on theoretical ideas that optimal engagement occurs when tasks are perceived as challenging but manageable (e.g. Brophy, 1999; Vygotsky, 1978) and on empirical findings suggesting that increases in self-efficacy do not linearly translate into better performance (e.g. Vancouver et al., 2008), we assume that the likelihood of engaging in physics-specific reasoning is highest within an intermediate range of MJOS values. If MJOS is too low, for example due to low expectancy or high perceived cost, students are unlikely to engage. If MJOS is excessively high, students may overconfidently rely on their initial interpretation without critically revising it, potentially leading to superficial or misguided reasoning. In this sense, the constitution of the MJOS right after initial task typification represents the point at which cognitive, metacognitive, and motivational processes converge to shape students' decision whether to engage in physics-specific reasoning for building a task representation or to proceed based on their initial task interpretation.

Having specified the assumptions underlying the TSMS model, we now summarise the proposed process logic in a semi-formal representation. This representation is not intended as a fully specified mathematical model, but as a concise clarification of the model's core relations. It makes explicit how the central constructs are assumed to relate to one another and provides the basis for deriving testable hypotheses.

## A Semi-Formal Representation of the TSMS Model

The following semi-formal representation specifies the TSMS model for physics tasks as the prototypical case developed in this paper. The central relations in the TSMS model can be represented semi-formally for student $i$ working on task $j$. In brief, the model can be represented as the following sequence:

$$Heuristic\ processes \rightarrow FIMM_{ij}, FOR_{ij}, FOD_{ij} \rightarrow SE_{ij}, V_{ij}, C_{ij} \rightarrow MJOS_{ij} \rightarrow P(E_{ij})$$

The construction of the first impression mental model of the task type, $FIMM_{ij}$, through heuristic processes is accompanied by a feeling of rightness $FOR_{ij}$ and a feeling of difficulty $FOD_{ij}$ which shape task-related self-efficacy $SE_{ij}$, value $V_{ij}$ and cost $C_{ij}$. The resulting motivationally informed judgement of solvability $MJOS_{ij}$ is assumed to predict the probability of physics-specific analytic engagement on the task $P(E_{ij})$.

$FOR_{ij}$ determines the extent to which task-related self-efficacy, value and cost are dominated by task-type-specific or general physics-related beliefs and attributions. We represent this by a weighting parameter $w_{ij}$, which increases monotonically with $FOR_{ij}$, such that $w_{ij} = s(FOR_{ij})$ and $0 \leq w_{ij} \leq 1$. For parsimony, we use the same weighting parameter for self-efficacy, value and cost, assuming that FOR reflects the perceived credibility of the initial task typification as a whole. $FOD_{ij}$ further specifies perceived cost. We model this by assuming that perceived cost increases with FOD, with a person-specific weighting parameter $\lambda_i > 0$ reflecting that students may differ in how strongly an initial feeling of difficulty translates into perceived cost:

$$SE_{ij} = w_{ij} SE_i^{type(j)} + (1 - w_{ij}) SE_i^{gen}$$

$$V_{ij} = w_{ij} V_i^{type(j)} + (1 - w_{ij}) V_i^{gen}$$

$$C_{ij} = w_{ij} C_i^{type(j)} + \left(1 - w_{ij}\right) C_i^{gen} + \lambda_i FOD_{ij}$$

In line with Eccles & Wigfield's (2020) situated expectancy-value theory and a formalisation proposed by Tzur et al. (2016), we define task-specific net value as value adjusted for perceived cost, allowing value and cost to differ in their relative contribution to net value:

$$NV_{ij} = \rho_{ij}V_{ij} - C_{ij}, \rho_{ij} > 0$$

While self-efficacy and net value should contribute independently to MJOS, empirical evidence supports the expectancy × value logic that self-efficacy becomes especially motivationally relevant when the perceived value of engaging with the task exceeds its perceived costs (Trautwein et al., 2012; Tzur et al., 2016). Therefore, $MJOS_{ij}$ is conceptualised as resulting from both additive and multiplicative effects of task-related self-efficacy and task-specific net value:

$$MJOS_{ij} = \alpha_0 + \alpha_1 SE_{ij} + \alpha_2 NV_{ij} + \alpha_3 (SE_{ij} \times NV_{ij}) + \epsilon_{ij}$$

Finally, physics-specific analytic engagement is expected to be most likely within an adaptive intermediate range of $MJOS_{ij}$: very low and very high $MJOS_{ij}$ values are expected to reduce the likelihood $P(E_{ij} = 1)$ of physics-specific analytic engagement. This can be represented by a negative quadratic effect:

$$P(E_{ij} = 1) = logit^{-1}(\beta_0 + \beta_1 MJOS_{ij} - \beta_2 MJOS_{ij}^2)$$

## Deriving Testable Hypotheses from the TSMS Model

Building on the process logic and the semi-formal representation of the TSMS model, the following hypotheses specify the model's core empirical implications for physics tasks.

**H1. FIMM-formation hypothesis:** During first task reading, physics task cues will elicit task-specific FIMMs that vary across tasks within and between students and guide students' initial response tendencies or solution approaches. H1 could be tested in a task sorting assessment by assessing students' sorting decisions and intuitive responses across tasks that systematically vary in surface and deep-structure cues.

**H2. Task-level specification hypothesis:** After initial task typification, motivational evaluations will vary across tasks within students to the extent that students experience their initial task typification as credible. Specifically, task-related self-efficacy, task-related value, and task-specific perceived cost should vary across tasks, even when students' general physics-related self-efficacy, value, and perceived cost are taken into account.

**H2a. FOR-specificity hypothesis:** The feeling of rightness associated with the FIMM will positively predict the extent to which self-efficacy, value and perceived cost become task-related after initial task typification. Higher FOR should be associated with stronger reliance on task-type-specific rather than more general self-efficacy, value and perceived cost.

**H2b. FOD-cost hypothesis:** The feeling of difficulty associated with the FIMM will positively predict task-specific perceived cost beyond general perceived cost and task-type-specific cost attributions. Higher FOD should therefore reduce task-specific net value, even when general perceived cost, task-type-specific cost attributions, and objective or expert-rated task difficulty are controlled. H2, H2a, and H2b could be tested in an experimental pre–post design with task-typification training. Before the training, students would report general physics-related self-efficacy, value, and perceived cost. After training on specific physics task types, task-type-specific self-efficacy, value, and cost attributions would be assessed for the trained types. Students would then work on matching and non-matching tasks and provide, before solving each task, an initial typification, FOR, FOD, and task-related ratings of self-efficacy, value, and perceived cost. H2 would be supported by within-student task-level variation beyond general evaluations; H2a by stronger alignment between task-related and task-type-specific evaluations when FOR is high; and H2b by an additional positive effect of FOD on task-specific perceived cost beyond general perceived cost, task-type-specific cost attributions, and expert-rated task difficulty. FOR and FOD could be assessed with brief rating items adapted from metacognitive judgement research (c.f. Efklides & Tsiora, 2002; Thompson et al., 2011).

**H3. MJOS-integration hypothesis:** MJOS will be predicted by task-related self-efficacy, task-specific net value, and their interaction. Specifically, self-efficacy and net value should contribute to MJOS, while the association between self-efficacy and MJOS should depend on net value: high positive net value should strengthen the positive effect of self-efficacy, whereas negative net value should weaken or even reverse it. H3 could be tested by assessing MJOS as a task-level judgement and modelling it as a function of task-related self-efficacy, task-specific net value, and their interaction. MJOS would need to be validated as a distinct task-level judgement. Task-specific judgement formats such as those used by Siefer et al. (2021) provide a promising starting point for developing such a measure.

**H4. Non-linear engagement-probability hypothesis:** The relation between MJOS and the probability of engagement in physics-specific reasoning processes for building a task representation will be non-linear. Specifically, $P(E_{ij} = 1)$ should be highest at intermediate MJOS values, whereas both very low and very high MJOS values should reduce this probability. For tasks that cue an incorrect or misleading FIMM, intermediate MJOS values should be associated with the highest probability that students revise their initial task representation. H4 could be examined by implementing a two-response paradigm in the design described above: students first provide an initial task typification or initial response under time constraints and then have the opportunity to revise these without such constraints (e.g., Thompson et al., 2011). Response revision, improved task typification, or deep-structure-based explanations could serve as indicators of analytic revision, and their probability should be highest at intermediate MJOS values. A key validity criterion for MJOS would be whether it predicts these indicators beyond task-related self-efficacy, value, and perceived cost considered separately.

# Discussion of the TSMS Model

In the first part of this paper, we introduced the TSMS model, justified its theoretical assumptions, provided a semi-formal representation of the model, and derived testable hypotheses for physics tasks. Although the present specification of the model is developed for physics tasks, we understand it as a prototypical case of a broader science-task problem. In the second part, we discuss the model's scope, boundaries, and limitations. We then examine its usefulness for interpreting existing research findings and outline ideas for future validation. Finally, we discuss the model's theoretical contribution and implications for research and practice.

## Scope, Boundaries and Limitations of the TSMS Model

The TSMS model provides a theoretical framework for understanding motivational and cognitive processes in the predecisional phase of task processing. In the present paper, the model is specified and formalised for physics tasks. It explains how the feeling of rightness (FOR) and feeling of difficulty (FOD) that accompany the first impression mental model of the task type (FIMM) can shape task-related expectancy-value evaluations. The resulting task-specific motivational stimulus, conceptualised as the motivationally informed judgement of solvability (MJOS), is assumed to influence the probability that students engage in disciplinary analytic reasoning to build or revise a task representation rather than relying on an initial interpretation that may be misleading. Thus, the TSMS model provides a theoretical basis for empirical investigations of early task engagement in physics and, potentially, in science tasks with similar characteristics.

### *Scope*

The TSMS model should be most informative for science tasks in which students must interpret surface features in relation to disciplinary concepts, models, representations, or practices before they can decide how to proceed. It should be particularly relevant when tasks can cue plausible but inappropriate initial interpretations, familiar but inadequate routines, or intuitive responses that require analytic reconsideration. In the physics-specific specification developed here, the model can be applied to any physics test tasks that require the activation of physics knowledge, such as knowledge-centred problems (c.f. Friege, 2001; Nguyen et al., 2020). However, it is most relevant for physics test tasks in which the underlying physics model or solution approach is not immediately transparent from surface features – and in particular for tasks that may trigger misleading FIMMs, non-physics-related FIMMs, or intuitive responses that require analytic reconsideration.

This includes for example, introductory mechanics tasks that trigger everyday intuitions or alternative conceptions, such as tasks used in prominent concept inventories like the Force Concept Inventory (Hestenes et al., 1992), but also more complex problem-solving tasks contextualised within socio-scientific issues that require multiperspective approaches, or more formal or advanced physics tasks in which equations, graphs, diagrams, or symbolic representations may cue inappropriate procedures or classical interpretations of non-classical phenomena (e.g. Baily & Finkelstein, 2010; Sherin, 2001; Tuminaro & Redish, 2007). Thus, the model is most informative whenever students must relate surface features to underlying science concepts, principles, or models before they can construct an adequate task representation.

***Boundaries***

The TSMS model describes the decision-making process through which students decide how to approach science tasks in subject-specific test situations. With minor modifications, the model could also be applicable to learning tasks. In this case, however, it is important to carefully incorporate goal orientations, such as mastery versus performance goal orientations, into the model because their relevance for cognitive engagement, achievement motivation, and achievement emotions in learning situations has been demonstrated in several theoretical models and empirical studies (cf. Pekrun, 2006; Wigfield & Eccles, 2000).

A further boundary concerns task presentation. The present physics-specific version of the TSMS model is primarily formulated for written tasks, because it focuses on first task reading and on the construction of an initial task representation from textual, symbolic, graphical, and contextual cues that remain available and can be revisited. Similar processes may occur in oral, experimental, or interactive settings, but cues such as a teacher's wording, gestures, and intonation, peer interactions among students, material objects, or experimental setups often unfold sequentially and may have more ephemeral mediality, making them less available for later revisiting (Kauertz et al., under review). Such settings may also be more strongly shaped by the local idioculture in which the task is reconstructed (Finkelstein, 2005). The model would therefore need to be adapted before being applied to these contexts.

Finally, the TSMS model focuses on the predecisional phase of task processing and does not explain the full problem-solving process or the likelihood of reaching a correct solution. Even if students make a serious attempt to solve a task and activate relevant knowledge or problem solving schemata, their ability, persistence, and effort may vary throughout the problem-solving process, and they may still give up or arrive at an incorrect solution. However, if students do not seriously engage with the task in a disciplinary way at the very beginning of task processing and instead rely on initial thoughts without activating analytic reasoning processes to build a task representation, they are more likely to adopt an inappropriate line of reasoning.

### *Limitations*

The TSMS model currently includes some black-box-simplifications. First, it does not yet specify in detail which person or task characteristics influence the process through which students decide how to engage with a task. In the physics-specific specification developed here, relevant person characteristics may include physics-related dispositions, such as everyday experiences, experiences in physics class or with a particular type of task, declarative, procedural, and conditional knowledge required to apply specific physics concepts (Chi et al., 1981), and beliefs about physics knowledge and learning (Gupta & Elby, 2011). More general abilities, such as cognitive reflection skills, may also play a role (Alinea, 2020; Frederick, 2005; Wood et al., 2016). On the task side, it may be useful to consider the complexity of both the surface and deep structure of the task (Löffler et al., 2018), as well as its representational format or context (e.g., real-life vs. physics context; Siefer et al., 2021).

Second, the model addressed how emotional engagement triggered in the situation frame might affect text comprehension processes and the construction of the FIMM, but it does not yet elaborate how emotions or mood may influence reliance on heuristic processing, although such effects have been reported (Park & Banaji, 2000). As such effects should manifest in the metacognitive experiences accompanying the FIMM, they may not threaten the validity of the model. However, this assumption requires empirical testing.

A further limitation concerns transferability to other science domains. Although the TSMS model is introduced as a model of task-specific engagement in science tasks, the present paper develops, visualises, formalises, and derives hypotheses for the model through the prototypical case of physics tasks. Applying the model to biology, chemistry, or other science domains would not necessarily require changing the model's general process logic, as visualised in Figure 1. Rather, the relevant disciplinary concepts, practices, representations, task types, motivational beliefs, and sources of misleading initial task typifications would have to be specified for the respective domain. Thus, Table 1, the semi-formal representation, and the hypotheses would need to be adapted primarily at the level of domain-specific constructs and operationalisations.

Finally, we want to stress that the TSMS model is not a synthesis of direct empirical evidence for all proposed relations, but rather provides theoretical grounds for future empirical research. Although many studies examined how expectation for success and value attribution contribute to achievement motivation, achievement emotions, and aspects of student engagement, to our knowledge, research in science education has rarely focused on motivational processes at the very beginning of task processing. The model therefore provides hypotheses and a semi-formal structure for investigating a decision process that remains empirically underexplored in science education.

**Usefulness and Future Validation of the TSMS Model**

Within the described boundaries, the usefulness of the TSMS model lies in its ability to interpret diverse findings from physics education research through a common mechanism: task cues may elicit a FIMM, the metacognitive experiences accompanying this FIMM may shape task-related self-efficacy, value, and perceived cost, and the resulting MJOS may influence whether students revise or rely on their initial interpretation. The following examples do not validate the model directly, but illustrate how it can organise existing findings and generate more specific empirical questions.

***Physics Tasks, Heuristic FIMMs, and Engagement***

Kryjevskaia et al. (2014) showed that students who had successfully applied relevant physics knowledge in screening items on capacitors nevertheless relied on misleading interpretations in target items. The authors explain this through heuristic reasoning based on the availability heuristic: highly accessible concepts such as conservation or equilibrium were activated by target item features and guided responses even when inappropriate. In TSMS terms, such concepts can become part of a misleading FIMM, linking the finding to the FIMM-formation hypothesis (H1). If this FIMM is accompanied by high FOR, students may rely on task-type-specific self-efficacy, value, and cost attributions associated with the cued interpretation. For example, the task may appear familiar, solvable, and not worth further effortful revision. Under these conditions, students may form a high MJOS and see little need to engage in physics-specific analytic reasoning. This is consistent with the FOR-specificity hypothesis (H2a) and the non-linear engagement-probability hypothesis (H4), especially the prediction that misleading FIMMs are most likely to be revised at intermediate MJOS values.

Speirs et al. (2021) show that information provided after students had selected answers in target items was often used to justify these answer unless it explicitly contradicted them. In TSMS terms, later information may be assimilated into the FIMM unless it lowers FOR. The apparent simplicity of the tasks may have supported a high MJOS because students may have felt that they had correctly identified the task type, perceived high chances of success, and expected low costs for proceeding with their initial interpretation. Under such conditions, analytic engagement would be less likely, because the initial interpretation appears both credible and sufficiently solvable. Contradictory information, however, may lower FOR and increase perceived difficulty, hence FOD. More general physics-related and potentially lower self-efficacy may then become more influential and higher perceived cost may reduce net value. Such changes may move MJOS away from an overconfidently high range toward an intermediate range in which analytic revision becomes more likely. This links the findings to the FOR-specificity hypothesis (H2a), the FOD-cost hypothesis (H2b), and the non-linear engagement-probability hypothesis (H4). Providing students with additional information beforehand (contradictory or not) may additionally provide evidence for the FIMM construction hypothesis (H1).

Heckler et al. (2010) showed that many physics students relied on an incorrect heuristic in graph tasks with misleading task cues, attending to height instead of slope, even though the tasks were very simple. Providing additional time and asking students to carefully reconsider their answer improved performance, but almost a third of students still maintained the incorrect response. In TSMS terms, the graph's salient surface feature may have cued a misleading FIMM, linking the finding to H1. Additional time may not help if this FIMM feels sufficiently right and if the task appears too simple to warrant revision. This is consistent with H2a and H4: high FOR and very high MJOS may reduce the perceived need for physics-specific analytic engagement.

Finally, Dufresne et al. (1992) showed that training students to classify and analyse problems according to deep structure can improve physics problem solving performance. In terms of the TSMS model, this can be interpreted as supporting the construction or revision of the FIMM by directing students' attention from surface features to relevant deep-structure features. Such training may therefore increase the likelihood of physics-specific analytic engagement and reduce the likelihood that students accept an inappropriate FIMM, linking the finding to H1 and H4.

### *Specification of Expectancy-value Evaluations Through FOR and FOD*

The TSMS model assumes that motivational evaluations become task-related through metacognitive experiences accompanying the FIMM. This concerns the task-level specification hypothesis (H2), the FOR-specificity hypothesis (H2a), and the FOD-cost hypothesis (H2b).

Direct evidence for this complete process in physics tasks is still lacking. However, Siefer et al. (2021) showed that students' self-efficacy judgements varied across linear function tasks with different characteristics, supporting the plausibility of task-level specification (H2). The TSMS model adds that such differences should arise when students typify a task and experience FOR and FOD in response to the resulting FIMM.

The model can also help explain heterogeneous findings on task-specific value. While Schukajlow et al. (2012) found no differences in students' values across mathematical problem types, Krawitz and Schukajlow (2018) found such differences and attribute discrepancies with earlier findings to differences in the samples, such as students' level of experience and academic track. The TSMS model provides a possible process explanation: if students do not clearly recognise a task type, FOR should be low and task-related value should remain dominated by general value attributions; if more experienced students recognise a task type with high FOR, task-type-specific value representations can be activated. This illustrates H2a.

With respect to FOD, metacognitive research suggests that experienced difficulty can inform estimates of required effort and regulate subsequent processing. Importantly, experienced difficulty does not necessarily reduce engagement: Alter et al. (2007) showed that metacognitive difficulty or disfluency can activate analytic reasoning and reduce reliance on intuitive heuristics. The TSMS model captures this dual role by assuming that FOD can contribute to perceived cost, while analytic engagement is most likely when MJOS remains within an adaptive intermediate range. This speaks to H2b and the non-linear engagement-probability hypothesis (H4).

### ***MJOS as Motivational Integration***

The MJOS-integration hypothesis (H3) states that task-related self-efficacy and task-specific net value contribute to MJOS both additively and interactively. This is consistent with Trautwein et al. (2012), who found an expectancy × value effect on achievement, and Tzur et al. (2016), who showed that the effect of self-efficacy on performance depends on reward. Crucially, MJOS is not assumed to directly predict performance, but the probability of physics-specific analytic engagement at the onset of task processing. Thus, the TSMS model does not merely restate expectancy–value theory. It explains how a physics task gives rise to a FIMM, how FOR and FOD specify motivational evaluations for that task, and how MJOS influences whether students validate or revise their initial interpretation.

Taken together, these reinterpretations illustrate how the TSMS model can organise multiple and partly heterogeneous findings through a shared mechanism. However, they do not constitute direct validation of the model. Direct validation requires empirical studies that assess FIMM, FOR, FOD, task-related self-efficacy, value, perceived cost, MJOS, and indicators of analytic revision within the same design.

## Theoretical Contribution and Practical Implications of the TSMS Model

In this paper, we provided a theoretical basis resulting in a testable model to describe and analyse how students decide to engage in a particular physics task. Focusing on the predecisional phase of task processing (Heckhausen & Gollwitzer, 1987) allowed us to integrate theories and models from educational psychology, cognitive psychology, and physics education research. We integrated elements of situated expectancy-value theory for achievement motivation (Eccles & Wigfield, 2020), control-value theory for achievement emotions (Pekrun, 2006; Pekrun et al., 2009; Pekrun & Linnenbrink-Garcia, 2012), heuristic-analytic theory of reasoning (Evans, 2006, 2019), and the metacognitive framework of reasoning (Thompson, 2009), and connected them to physics education research on problem solving, problem representation, expert–novice differences, and resources-based accounts of students' reasoning (e.g. Chi et al., 1981; Hammer et al., 2005; Heckler, 2011; Löffler et al., 2018; Tuminaro & Redish, 2007). This approach enables us to address several desiderata formulated by researchers in these fields, which we will shortly address in the following paragraphs. Finally, we will address in more detail how the TSMS model can contribute to research on science-specific aspects of engagement and on commonsense physics, as well as provide implications for science education research and practice.

### *Contribution to research on expectancy-value theory*

The TSMS model contributes to expectancy–value theory by proposing a theoretical mechanism for how expectancy–value evaluations may become task-related at the very beginning of task processing. This responds to situated expectancy–value theory which emphasise context-sensitivity of expectancy and value beliefs and Eccles and Wigfield (2020) call for more research on the "sources of information individuals […] use, the processes by which they form their ASCs [Academic Self Concepts] and STVs [Subjective Task Values], and the nature of the specific hierarchies of ASCs and STVs being activated for any specific achievement-related choice" (p. 1). While many empirical studies still operationalise expectancy at broader test-, course-, or domain-levels (e.g., Asseburg & Frey, 2013; Frenzel et al., 2007; Wigfield & Cambria, 2010), the TSMS model uses heuristic–analytic theory of reasoning to explain how initial task typification, FIMM, FOR, and FOD shape task-related self-efficacy, value, perceived cost, and MJOS. Through its semi-formal representation, the model also provides a starting point for more fine-grained empirical studies of task-level expectancy–value evaluations.

***Contribution to research on dual process theories of reasoning***

The TSMS model contributes to research on dual-process theories of reasoning by proposing a theoretical mechanism for how analytic revision of an initial task interpretation may be triggered in authentic physics tasks. This responds to Pennycook's (2018) observation that one, 'of the most important gaps [in dual process theory literature] is the lack of solid explanation of how analytic thinking is actually triggered' (p.20). Whereas many dual-process reasoning tasks have been criticised for their artificiality because they are deliberately designed to contain misleading cues (cf. Stanovich, 2018), physics tasks can naturally involve similar conflicts, as salient surface features, everyday experiences, familiar representations, or routine solution strategies may cue plausible but inappropriate initial interpretations. By linking heuristic–analytic reasoning with physics problem solving and expectancy-value theory, the TSMS model explains how FIMM, FOR, FOD, and MJOS shape whether students revise, elaborate, justify, or simply proceed with an initial task interpretation and provides a theoretical basis for analysing dual-process phenomena in non-artificial physics tasks.

***Contribution to research on science-specific aspects of engagement and on commonsense physics***

The TSMS model responds to Sinatra et al.’s (2015) call for more focused theoretical work on the domain-specific aspects of engagement. It specifies physics-specific cognitive-behavioural engagement at the beginning of task processing as physics-specific reasoning processes for building a task representation by questioning an initial task interpretation in light of relevant physics concepts, principles, and models. Thus, the model does not treat every form of strategy use or self-regulation as physics-specific engagement: students may apply a formula, use an analogy, or search for a familiar answer without questioning whether this approach fits the task. In addition, the model describes how cognitive, affective, and motivational dispositions may interact in shaping whether students engage with a task in a physics-specific way.

The TSMS model may also contribute to the ongoing debate in research on commonsense or intuitive physics. Researchers adopting a 'coherentist' perspective argue that misconception-like answers stem from students holding theory-like alternative conceptions about how the world operates, which can coexist alongside scientific concepts. In contrast, researchers with a 'fragmentist' viewpoint assume that novice students’ knowledge is fragmented into 'pieces of knowledge' or 'resources'. These fragments are activated by the contextual characteristics of a task (diSessa, 1993, 2014; Hammer et al., 2005; Sabella & Redish, 2007). Consequently, they posit that misconception-like responses result from reasoning based on incorrectly connected fragments rather than from coherent alternative conceptions. The TSMS model offers a potential solution to integrate these conflicting perspectives by following an idea originally proposed by Badagnani et al. (2017). When students read a physics task, they build a first impression mental model (FIMM) of the task type which describes the structure of the most plausible resources activated by the task description. If the motivationally informed judgement of solvability (MJOS) is very low or very high, the probability that students initiate analytic reasoning processes to question their FIMM by consciously activating conceptual knowledge for building a task representation is rather low, and the probability that they just choose the answer option that corresponds best to their FIMM is quite high. However, if they are asked to justify their answer, they must activate some kind of conceptual knowledge. If a student holds strong alternative conceptions (e.g., impetus theory) parallel to physics concepts, based on the availability heuristic, the former might be activated more easily and, therefore, might be used to justify the intuitive answer. This would explain why students refer to coherent alternative conceptions when asked to justify wrong answers on specific items (e.g., Bekaert et al., 2022; Hestenes et al., 1992), although their answer patterns over all items are quite incoherent regarding such alternative conceptions (e.g., Heller & Huffman, 1995; Lasry et al., 2011; Stewart et al., 2007).

### *Implications of the TSMS model for science education research and practice*

Empirical research in science education shows that students' reasoning in tasks can be strongly influenced by situational and contextual factors and that interactions between students' cognitive, affective, and motivational dispositions can influence the interpretation and validity of test performance measures. Future research can build on the TSMS model by empirically testing it and using it to quantify the influence of motivational and metacognitive processes on task performance in science tasks. The model also offers a theoretical basis for designing adaptive multiple-choice test instruments to distinguish between missing conceptual or procedural knowledge and a failure to activate science-specific reasoning. For example, the two-response paradigm used in dual-process research (e.g., Thompson et al., 2011) could help distinguish unexamined or merely justified initial responses from responses revised through disciplinary analytic processing. Additional prompts for students who do not revise incorrect initial responses (see Talanquer, 2017 for an example for such prompts in chemistry education) could then help clarify whether their difficulty lies in missing knowledge or in an initial task interpretation that was not critically examined.

The TSMS model suggests that teaching students about task typification explicitly could help them to better understand the tasks, to identify necessary knowledge, and apply appropriate skills. Such instruction may support students in constructing more adequate FIMMs and in experiencing sufficient FOR to specify task-related motivational evaluations on the basis of the task at hand rather than relying primarily on general self-efficacy, value, or cost beliefs. Since teachers in science classes often differentiate between students by assigning different tasks, understanding task typification, engagement, and knowledge activation processes could inform the design of curricula, tasks, and scaffolds that help students relate surface features to underlying science concepts, principles, and models. Such insights may also inform research on students' conceptions by explaining why students sometimes give misconception-like answers instead of activating previously learned science concepts. Moreover, by specifying underlying motivational processes, the model may guide the search for post hoc explanations of unusual or unexpected variance in all kinds of task-based competence measures.

## Conclusion

Students' early engagement with science tasks is not determined by knowledge and skills alone. Disengaging from tasks is a common problem in science education, where task solving plays an important role in learning, diagnosis, and assessment. Across the natural sciences, task solving often requires students to go beyond everyday knowledge and engage in cognitive modelling, formalisation, rational reasoning, and evidence-based argumentation. At the same time, science tasks may activate intuitive ideas about how the world works that conflict with disciplinary concepts, models, or procedures. Dealing with such conflicts requires not only knowledge and skills, but also the willingness to invest effort and engage in disciplinary analytic reasoning.

The TSMS model provides a theoretical account of this early decision process. Drawing on established models from cognitive psychology, motivational theory, and physics education research, it describes possible mechanisms through which initial task interpretations, metacognitive experiences, and task-related motivational evaluations shape students' engagement with a physics task. These mechanisms were translated into a semi-formal representation and testable hypotheses using the simplest mathematical relations that are consistent with the theoretical assumptions. Although the present paper develops and formalises the model for physics tasks, we understand this specification as a prototypical case. Applications to other science domains would require specifying the relevant disciplinary concepts, representations, practices, task types, and sources of misleading initial typifications for the respective domain.

Overall, the TSMS model aims to advance research in science education by making the predecisional phase of task processing theoretically visible and empirically accessible. We hope to show that this account can provide a useful tool for addressing open questions and critiques in different fields of research, for informing task, instruction, and assessment designs, and for advancing our understanding of cognitive and motivational processes relevant to science teaching and learning. Therefore, we invite researchers from relevant areas to critique, test, and refine the TSMS model.